\documentclass[journal=jctcce,manuscript=article,hyperref,layout=twocolumn]{achemso}
\setkeys{acs}{doi=true}
\setkeys{acs}{chaptertitle=true}
\usepackage[version=3]{mhchem} 
\usepackage[table,xcdraw]{xcolor}
\usepackage{comment}
\usepackage{threeparttable}
\usepackage{booktabs}
\usepackage{siunitx}
\usepackage{multirow}
\usepackage{amsmath}
\usepackage{orcidlink}
\usepackage{graphicx}
\usepackage{url}
\makeatletter
\newcommand*\doi{\begingroup\let\do\@makeother\dospecials
  \catcode`\{=1 \catcode`\}=2 \doi@aux}
\newcommand*\doi@aux[1]{\endgroup\href{https://doi.org/#1}{\nolinkurl{#1}}}
\makeatother

\newcommand{\kcalmol}{\mbox{kcal$\cdot$mol$^{-1}$}\,}

\author{Jan M.L. Martin\orcidlink{0000-0002-0005-5074}}
\email{gershom@weizmann.ac.il}
\phone{+972 8 9342533}
\fax{+972 8 9343029}
\affiliation{Department of Molecular Chemistry and Materials Science, Weizmann Institute of Science, 7610001 Re\d{h}ovot, Israel}    
\date{\today} 

\title[post-CCSDT) extrap.]{Basis set extrapolation for post-CCSD(T) contributions reconsidered}
\abbreviations{CCSDT, CCSDT-3, CCSDT(Q), CCSDT(Q)$_\Lambda$}
\begin{document}
\maketitle

\begin{center}
\small Preprint \today
\end{center}

\begin{abstract}
Higher-order correlation contributions beyond CCSD(T) are a key ingredient of high-accuracy computational thermochemistry and spectroscopy approaches. In an attempt to more firmly rationalize their basis set convergence behavior, we have carried out benchmark studies for over three dozen molecules with basis sets through cc-pV7Z. Connected quadruples (Q) converge smoothly to the basis set limit if evaluated for at least cc-pVTZ basis sets; cc-pVDZ is anomalous for ionic systems. Post-CCSDT(Q) contributions converge quite rapidly with the basis set, with even a truncated cc-pVTZ(d) basis set yielding results near the basis set limit. In contrast, higher-order triples contributions, $T_3-(T)$, converge slowly and at times erratically: we show that this is the superposition of three different convergence behaviors, namely (a) CCSDT-3 $-$ CCSD(T) with leading term $E^{(5)}_{TQ}$ (moderately fast); (b) CCSDT $-$ CCSDT-4 (higher-order triples coupling terms, very fast); and (c) triples-on-triples ladder diagrams (slow) in CCSDT-4 $-$ CCSDT-3, with leading term $E^{(5)}_{TT}$. The slow convergence of the latter dominates for large basis sets. 
A Petersson-style $(L+a)^{-3}$ extrapolation well describes convergence behavior with different shifts $a=+1.10$ for CCSDT-3 $-$ CCSD(T), $a=+0.84$ for CCSDT $-$ CCSDT-3, $a=+0.69$ for combined CCSDT $-$ CCSD(T), and $a=-0.66$ for (Q). 
\end{abstract}

\section{Introduction}

Computational thermochemistry has reached an accuracy level that is competitive for most species with the best available experimental data, such as those compiled in the Active Thermochemical Tables (ATcT) project of Ruscic and co-workers.(Ref.\cite{jmlm345} and references therein.)

While CCSD(T) has a well-deserved status as “the gold standard of quantum chemistry” (Thom H. Dunning Jr.), it is well known to be insufficient for kJ/mol or better accuracy in thermochemistry, or  even \kcalmol accuracy for molecules with some static correlation.

Hence, the two major families of high-accuracy computational chemistry methods both include post-CCSD(T) corrections. These families are HEAT (High Accuracy Extrapolated Ab Initio Thermochemistry\cite{HEAT,HEAT2,HEAT3,HEAT4}) of the late lamented John F. Stanton and co-workers, and the W$n$, or Weizmann-$n$, family of Martin and co-workers.\cite{jmlm127,jmlm200,jmlm205,jmlm235,jmlm269,jmlm273}

In the latest incarnation of the latter, W5preview1 by Barman et al.,\cite{jmlm340} we even include subvalence correlation beyond the CCSD(T) level, as does the super-HEAT project of Thorpe et al.\cite{superHEAT}

But while there is a significant literature (e.g., Refs.\cite{KutzelniggMorgan1992Rates,jmlm096,Halkier1998a,Klopper2001,Crawford2006,ranasinghe2013ccsd,IronBloke} and references therein) on basis set extrapolation of the CCSD and CCSD(T) correlation energy to the one-particle basis set limit, the literature on post-CCSD(T) corrections is quite limited. The most representative paper is the 2020 study by Karton,\cite{Karton2020} which proposes two-point extrapolation exponents for CCSDT $-$ CCSD(T), (Q), and CCSDTQ $-$ CCSDT(Q), obtained from fitting against values at the quintuple- and sextuple-zeta basis set level for a baker's dozen of molecules. We actually adopted this work in the W5preview paper,\cite{jmlm340} while in other thermochemistry papers, more ad hoc procedures were used.

These corrections are not trivial academic matters. For a molecule like ozone, they reach several \kcalmol. The importance of these corrections is strongly linked to their non-dynamical correlation character, for which a plethora of diagnostics has been developed, most recently the DAD or density asymmetry diagnostic of Stanton and co-workers,\cite{jmlm337} and a natural orbital triples increment-based diagnostic by Jones et al.\cite{jmlm339} Reviews of some of the other commonly used diagnostics are given by Fogueri et al.\cite{jmlm244} and by Matito and co-workers.\cite{Matito2024,Matito2025}

Recent advances in electronic structure software, particularly accelerations in the CFOUR electronic structure program system,\cite{CFOUR,CFOURcode} and specifically in its NCC or new coupled cluster module,\cite{NCC1,NCC2,TBLIS} have enabled coupled cluster calculations on commodity hardware that were hitherto beyond the reach of even well-equipped research groups.

In the present paper, we will present reference data going all the way to septuple-zeta for an adequately large and chemically diverse sample. We will analyze the convergence behavior at each step in the coupled cluster hierarchy in detail, and we will propose revised extrapolations for the various terms.

We will show that the comparatively slow convergence behavior of CCSDT $-$ CCSD(T) actually results from the ``$\hat{T}_3$-on-$\hat{T}_3$'' term, which starts at fifth order\cite{Cremer2001} and which corresponds to the CCSDT-4\cite{CCSDT-4andCCSDT} $-$ CCSDT-3\cite{CCSDT-3} difference. We will also show that as one works further up the coupled cluster hierarchy, radial basis set convergence becomes increasingly more important than angular correlation convergence, and hence angular momentum truncated basis sets become an attractive option. Finally, we will show that the Petersson formula\cite{ranasinghe2013ccsd} with a shift, $(L+a)^{-3}$, is a very serviceable extrapolation formula, and that the shift $a$ is specific to the particular component we are investigating.

\section{Computational Details}
\label{sec:methods}

All calculations, including CCSD(T),\cite{Raghavachari1989,Watts1993} CCSDT-2\cite{CCSDT-3},
CCSDT-3\cite{CCSDT-3}, CCSDT-4\cite{CCSDT-4andCCSDT},  CCSDT\cite{Urban1985,CCSDT-4andCCSDT}, 
CCSDT(Q) Bomble et al.\cite{mrcc8,mrcc63,mrcc65} and Lambda coupled cluster\cite{lambdastanton1,lambdastanton2,lambdabartlett1,lambdabartlett2}
were performed using a development version of CFOUR\cite{CFOUR,CFOURcode} running on Intel Ice Lake (52 cores, 768 GB RAM) and AMD Zen4 (256 cores, 1.5 TB RAM) nodes of the Faculty of Chemistry's HPC facility CHEMFARM. 

The said development version includes several enhancements by Dr. Gregory H. Jones (U. of Florida Quantum Theory Project, presently at MIT). These include an open-shell CCSDTQ implementation documented in Ref.\cite{jmlm343}, as well as a modification (previously reported in Ref.\cite{jmlm340}) to the CCSDT(Q) and CCSDT(Q)$_\lambda$ code in NCC, in which the perturbation theory denominators are precomputed in an array rather than evaluated on the fly. The latter option is more parsimonious in memory and allows for nonstandard perturbation partitionings, but precomputation drastically enhances parallelization.

Basis sets considered were of the correlation consistent family. They include the original cc-pV$n$Z basis sets\cite{dunning1989gaussian,dunning1996gaussian6} as well as the diffuse-function augmented variants, where haV$n$Z stands for the combination of cc-pVnZ on hydrogen with aug-cc-pVnZ\cite{Kendall1992} on other elements. (This notation is due to Pavel Hobza; alternative notations are aug$'$-cc-pVnZ due to Janet E. Del Bene and jul-cc-pVnZ due to Papajak and Truhlar.\cite{calendar_basis}) 
Basis sets, including cc-pV7Z (Ref.\cite{Feller2011} and references therein), were downloaded from the ccRepo repository.
\url{http://www.grant-hill.group.shef.ac.uk/ccrepo/}.
We also considered several truncated versions of these basis sets, in an attempt to separate radial from angular flexibility. Finally, the cc-pwCVTZ and cc-pwCVQZ core-valence weighted basis sets\cite{Peterson2002} were also investigated, despite all calculations in this paper being frozen-core.

A 49-system subset of the W4-17 thermochemical benchmark\cite{jmlm273} was used as the primary test set. Valence-correlated CCSD(T)/cc-pV(Q+d)Z reference geometries were taken as is from the ESI of Ref.\cite{jmlm273} and not optimized further. UHF references were used throughout for open-shell systems. For the avoidance of doubt, all thermochemical discussion in this paper concerns TAEs (total atomization energies) in \kcalmol, hence a positive sign means a bonding contribution and a negative sign an antibonding one. Extrapolation from two basis sets is indicated by the `brace notation': e.g., cc-pV\{6,7\}Z indicates extrapolation from cc-pV6Z and cc-pV7Z basis sets.

For particularly slow calculations, estimated energies were monitored during iterations using the Aitken $\Delta^2$ procedure\cite{Aitken1926Delta2} implemented in a Python script supplied in the ESI. Particularly for NCC, the estimates from even a modest number of iterations were found to agree surprisingly well  with the final converged results.

\section{Results and Discussion}

Complete data are available as comma-separated value files in the Supporting Information, together with Python scripts that allow generating the atomization energies and the statistics. Furthermore, a full table of all the post-CCSD(T) contributions for all molecules can be found in the Supporting Information as well.

\begin{comment}\subsection{Splice this in somewhere}

The slower basis set convergence of ladder-type diagrams, compared to ring diagrams, is discussed by Irmler et al.\cite{Irmler2019PRL,Irmler2019JCP,Irmler2024}
\end{comment}

\subsection{Basis set convergence: higher-order triples}

\begin{comment}
\begin{table*}
  \caption{Basis set convergence (\kcalmol) of post-CCSD(T) contributions to the TAE
    for the six species with the largest $[(\textrm{Q})]$ and
    $[\textrm{CCSDT}]-[\textrm{CCSD(T)}]$; valence electrons correlated.}
  \label{tab:topsix}
  \input{tables/top-six}
\end{table*}
\end{comment}

\begin{table*}[!t]
  \caption{Basis-set convergence of the triples contributions to TAE (\kcalmol) for seven difficult species; valence electrons correlated.}
  \label{tab:seven_triples}
  \footnotesize
\begin{tabular*}{\textwidth}{@{\extracolsep{\fill}}l*{9}{S[table-format=-1.3]}@{}}
\toprule
 & {BN} & {B$_2$} & {C$_2$} & {P$_2$} & {N$_2$O} & {O$_3$} & {NO$_2$} & {CN} & {CCH} \\
\midrule
\multicolumn{10}{@{}l}{$[\textrm{CCSDT-3}]-[\textrm{CCSD(T)}]$~~~~~~\textbf{leading $E^{(5)}_{TQ_{dd}}$ , moderately fast, basis set slope positive}} \\
VTZ & -0.712 & -0.971 & -2.297 & -0.709 & -0.270 & -0.903 & 0.070 & 0.127 & -0.174 \\
VQZ & -0.577 & -0.937 & -2.191 & -0.677 & -0.208 & -0.801 & 0.133 & 0.176 & -0.136 \\
V5Z & -0.518 & -0.923 & -2.147 & -0.642 & -0.170 & -0.739 & 0.175 & 0.205 & -0.116 \\
V6Z & -0.491 & -0.915 & -2.127 & -0.625 & -0.149 & -0.705 & 0.198 & 0.219 & -0.106 \\
V7Z & -0.476 & -0.911 & -2.116 & -0.614 & -0.136 & -0.684 & 0.212 & 0.228 & -0.101 \\
REF2$^{a}$ & -0.449 & -0.903 & -2.096 & -0.599 & -0.116 & -0.652 & 0.234 & 0.241 & -0.090 \\
REF1$^{b}$ & -0.448 & -0.903 & -2.095 & -0.593 & -0.112 & -0.645 & 0.238 & 0.245 & -0.092 \\
\midrule
\multicolumn{10}{@{}l}{$[\textrm{CCSDT-3}]-[\textrm{CCSDT-2}]$~~~~~~\textbf{$T_1$ and $T_1T_2$ terms in the $T_3$ equation, fast, basis set slope positive}} \\
VTZ & 0.011 & 0.062 & 0.125 & 0.082 & 0.324 & 0.379 & 0.344 & 1.104 & 0.485 \\
VQZ & 0.053 & 0.072 & 0.141 & 0.083 & 0.325 & 0.414 & 0.361 & 1.152 & 0.511 \\
V5Z & 0.064 & 0.076 & 0.146 & 0.084 & 0.325 & 0.431 & 0.369 & 1.170 & 0.521 \\
V6Z & 0.067 & 0.077 & 0.147 & 0.084 & 0.325 & 0.437 & 0.374 & 1.176 & 0.524 \\
\midrule
\multicolumn{10}{@{}l}{$[\textrm{CCSDT-4}]-[\textrm{CCSDT-3}]$~~~\textbf{ leading $E^{(5)}_{TT}$, $W.T_3$, very slow, basis set slope negative}} \\
VTZ & 0.023 & 2.808 & 2.181 & 0.867 & 0.143 & 2.650 & 0.727 & 2.071 & 1.160 \\
VQZ & -0.166 & 2.659 & 1.930 & 0.771 & -0.038 & 2.452 & 0.559 & 1.968 & 1.050 \\
V5Z & -0.267 & 2.582 & 1.811 & 0.708 & -0.137 & 2.326 & 0.459 & 1.897 & 0.992 \\
V6Z & -0.318 & 2.544 & 1.752 & 0.662 & -0.188 & 2.261 &0.411 & 1.857 & 0.961 \\
%REF2b$^{c}$ & -0.397 & 2.485 & 1.660 & 0.590 & {---} & {---} & {---} & 1.795 & 0.913 \\
\midrule
\multicolumn{10}{@{}l}{$[\textrm{CCSDT}]-[\textrm{CCSDT-4}]$~~~~~\textbf{$W.T_1T_3+W.T_2T_3+...$, quite fast, basis set slope negative}} \\
VTZ & -1.757 & -1.478 & -1.796 & -0.880 & -1.074 & -2.740 & -1.374 & -1.214 & -0.630 \\
VQZ & -1.812 & -1.505 & -1.845 & -0.936 & -1.137 & -2.875 & -1.459 & -1.275 & -0.663 \\
V5Z & -1.826 & -1.510 & -1.858 & -0.958 & -1.156 & -2.925 & -1.485 & -1.291 & -0.671 \\
V6Z & -1.830 & -1.510 & -1.861 & -0.964 & -1.162 & -2.942 & -1.496 & -1.295 & -0.673 \\
%REF2b$^{c}$ & -1.836 & -1.510 & -1.866 & -0.973 & {---} & {---} & {---} & -1.301 & -0.676 \\
\midrule
\multicolumn{10}{@{}l}{$[\textrm{CCSDT}]-[\textrm{CCSDT-3}]$~~~~~~\textbf{dominated by $W.T_3$ for large basis sets, basis set slope negative}} \\
VTZ & -1.734 & 1.330 & 0.385 & -0.013 & -0.932 & -0.090 & -0.648 & 0.857 & 0.530 \\
VQZ & -1.978 & 1.154 & 0.085 & -0.166 & -1.175 & -0.423 & -0.900 & 0.693 & 0.387 \\
V5Z & -2.093 & 1.072 & -0.047 & -0.249 & -1.293 & -0.598 & -1.026 & 0.606 & 0.321 \\
V6Z & -2.148 & 1.034 & -0.108 & -0.302 & -1.350 & -0.681 & -1.085 & 0.562 & 0.287 \\
V7Z & -2.178 & 1.013 & -0.142 & -0.333 & -1.382 & -0.727 & -1.118 & 0.536 & 0.269 \\
REF2$^{a}$ & -2.234 & 0.975 & -0.203 & -0.384 & -1.439 & -0.810 & -1.177 & 0.494 & 0.234 \\
REF1$^{b}$ & -2.234 & 0.974 & -0.206 & -0.391 & -1.442 & -0.814 & -1.180 & 0.487 & 0.235 \\
\midrule
\multicolumn{10}{@{}l}{$[\textrm{CCSDT}]-[\textrm{CCSD(T)}]$~~~~~~\textbf{somewhat erratic for smaller basis sets, slope negative}} \\
VTZ & -2.446 & 0.359 & -1.912 & -0.722 & -1.202 & -0.993 & -0.578 & 0.984 & 0.356 \\
VQZ & -2.555 & 0.217 & -2.106 & -0.843 & -1.383 & -1.224 & -0.767 & 0.869 & 0.251 \\
V5Z & -2.611 & 0.149 & -2.194 & -0.891 & -1.463 & -1.337 & -0.851 & 0.811 & 0.205 \\
V6Z & -2.639 & 0.119 & -2.235 & -0.927 & -1.499 & -1.386 & -0.887 & 0.781 & 0.181 \\
V7Z & -2.654 & 0.102 & -2.258 & -0.947 & -1.518 & -1.411 & -0.906 & 0.764 & 0.168 \\
REF2$^{a}$ & -2.683 & 0.072 & -2.299 & -0.983 & -1.555 & -1.462 & -0.943 & 0.734 & 0.144 \\
REF1$^{b}$ & -2.682 & 0.070 & -2.301 & -0.985 & -1.554 & -1.458 & -0.942 & 0.732 & 0.144 \\
\bottomrule
\end{tabular*}

\smallskip
\parbox{\textwidth}{\footnotesize
V$n$Z = cc-pV$n$Z. $^{a}$\,REF2: Riemann3 extrapolation from cc-pV\{5,6\}Z, $E_\infty = E_6 + 1.5548\,(E_6-E_5)$. $^{b}$\,REF1: Riemann3 extrapolation from cc-pV\{6,7\}Z, $E_\infty = E_7 + 1.8805\,(E_7-E_6)$.}
\end{table*}

\begin{table*}[!t]
  \caption{Basis-set convergence of the quadruples contributions to TAE (\kcalmol) for seven difficult species; valence electrons correlated.}
  \label{tab:seven_quads}
  \footnotesize
\begin{tabular*}{\textwidth}{@{\extracolsep{\fill}}l*{7}{S[table-format=-1.3]}@{}}
\toprule
 & {BN} & {B$_2$} & {C$_2$} & {P$_2$} & {N$_2$O} & {O$_3$} & {NO$_2$} \\
\midrule
\multicolumn{8}{@{}l}{$[(\textrm{Q})]=[\textrm{CCSDT(Q)}]-[\textrm{CCSDT}]$} \\
VTZ(d)$^{a}$ & 2.911 & 1.081 & 3.014 & 1.186 & 2.160 & 4.214 & 2.022 \\
VTZ & 3.026 & 1.163 & 3.220 & 1.431 & 2.259 & 4.383 & 2.123 \\
VQZ(f)$^{a}$ & 3.138 & 1.205 & 3.326 & 1.509 & 2.358 & 4.587 & 2.240 \\
VQZ & 3.173 & 1.220 & 3.352 & 1.567 & 2.380 & 4.621 & 2.261 \\
V5Z & 3.223 & 1.239 & 3.393 & 1.608 & 2.418 & 4.713 & 2.310 \\
V6Z & 3.240 & 1.245 & 3.407 & 1.625 & 2.430 & 4.747 & 2.328 \\
V7Z & 3.248 & 1.248 & 3.413 & 1.633 & {---} & {---} & {---} \\
REF2$^{b}$ & 3.266 & 1.254 & 3.429 & 1.651 & 2.449 & 4.800 & 2.356 \\
REF1$^{c}$ & 3.263 & 1.254 & 3.424 & 1.648 & {---} & {---} & {---} \\
\midrule
\multicolumn{8}{@{}l}{$[(\textrm{Q})_\Lambda]-[(\textrm{Q})]$} \\
VDZ & -0.722 & -0.202 & -0.490 & -0.064 & -0.159 & -0.495 & -0.160 \\
VTZ(d)$^{a}$ & -0.872 & -0.243 & -0.549 & -0.073 & -0.143 & -0.498 & -0.164 \\
VTZ & -0.843 & -0.258 & -0.572 & -0.081 & -0.145 & -0.514 & -0.168 \\
VQZ(f)$^{a}$ & -0.867 & -0.266 & -0.587 & -0.083 & -0.142 & -0.529 & -0.170 \\
VQZ & -0.866 & -0.270 & -0.591 & -0.085 & -0.140 & -0.530 & -0.168 \\
V5Z(f)$^{a}$ & -0.870 & -0.269 & -0.591 & -0.083 & -0.139 & -0.535 & -0.171 \\
REF3$^{d}$ & -0.869 & -0.273 & -0.595 & -0.084 & -0.137 & -0.535 & -0.168 \\
\midrule
\multicolumn{8}{@{}l}{$[\textrm{CCSDTQ}]-[\textrm{CCSDT(Q)}_\Lambda]$} \\
VDZ & -0.365 & 0.286 & -0.578 & -0.044 & -0.296 & -0.425 & -0.232 \\
VTZ(d)$^{a}$ & -0.339 & 0.293 & -0.570 & -0.061 & -0.298 & -0.463 & -0.236 \\
VTZ & -0.312 & 0.289 & -0.530 & -0.065 & -0.299 & -0.464 & -0.230 \\
VQZ(f)$^{a}$ & -0.316 & 0.285 & -0.534 & -0.074 & -0.319 & -0.508 & {---} \\
VQZ & -0.316 & 0.279 & -0.537 & -0.084 & -0.324 & -0.516 & {---} \\
REF4$^{e}$ & -0.319 & 0.282 & -0.537 & -0.076 & {---} & {---} & {---} \\
\bottomrule
\end{tabular*}

\smallskip
\parbox{\textwidth}{\footnotesize
V$n$Z = cc-pV$n$Z. $^{a}$\,V$n$Z($l$): cc-pV$n$Z truncated at angular momentum $l$ (hydrogen one $l$ lower). $^{b}$\,REF2: Riemann3 extrapolation from cc-pV\{5,6\}Z, $E_\infty = E_6 + 1.5548\,(E_6-E_5)$. $^{c}$\,REF1: Riemann3 extrapolation from cc-pV\{6,7\}Z, $E_\infty = E_7 + 1.8805\,(E_7-E_6)$. $^{d}$\,REF3 = cc-pV5Z(g). $^{e}$\,REF4 = cc-pV5Z(f). 
%$^{f}$\,Rough estimate from Aitken extrapolation of first 5 iterations, corrected for error in same extrapolation for cc-pVQZ(f).
---: not available or not presently feasible.}
\end{table*}

\begin{table*}[!t]
  \caption{Largest absolute $T_1$, $T_2$, $T_3$, and $T_4$ amplitudes in valence CCSDTQ/cc-pVTZ wavefunctions. Degeneracies, where greater than one, are given as superscripts in parentheses.}
  \label{tab:maxamp}
  \footnotesize
\setlength{\tabcolsep}{3pt}
\begin{tabular}[t]{@{}l*{4}{r@{\hspace{17pt}}}@{}}
\toprule
Species & {$\max|T_1|$} & {$\max|T_2|$} & {$\max|T_3|$} & {$\max|T_4|$} \\
\midrule
C$_2$ & 0.08508 & 0.38115 & 0.00869\rlap{$^{(2)}$} & 0.00138 \\
B$_2$ & 0.07505 & 0.34582 & 0.02518\rlap{$^{(2)}$} & 0.01455 \\
BN & 0.20411 & 0.32637 & 0.02783\rlap{$^{(2)}$} & 0.00327\rlap{$^{(4)}$} \\
O$_3$ & 0.08492 & 0.29868 & 0.01334 & 0.00051 \\
F$_2$ & 0.01661 & 0.19456 & 0.00200\rlap{$^{(2)}$} & 0.00035 \\
P$_2$ & 0.03301\rlap{$^{(2)}$} & 0.14325\rlap{$^{(2)}$} & 0.00594\rlap{$^{(2)}$} & 0.00088\rlap{$^{(2)}$} \\
NO & 0.09086 & 0.12956 & 0.02016 & 0.00100 \\
NO$_2$ & 0.06719 & 0.12531 & 0.02827 & 0.00204 \\
O$_2$ & 0.03453\rlap{$^{(2)}$} & 0.11674 & 0.00462\rlap{$^{(2)}$} & 0.00165 \\
AlF & 0.02346\rlap{$^{(2)}$} & 0.11347\rlap{$^{(2)}$} & 0.00128\rlap{$^{(2)}$} & 0.00015\rlap{$^{(2)}$} \\
OF & 0.13672 & 0.11313 & 0.02137 & 0.00133 \\
BH & 0.01755 & 0.11243\rlap{$^{(2)}$} & 0.00221\rlap{$^{(2)}$} & 0.00036\rlap{$^{(2)}$} \\
S$_2$ & 0.03942\rlap{$^{(2)}$} & 0.11083 & 0.00715\rlap{$^{(2)}$} & 0.00090 \\
Cl$_2$ & 0.01748 & 0.11005 & 0.00198\rlap{$^{(2)}$} & 0.00008\rlap{$^{(2)}$} \\
CS & 0.03370\rlap{$^{(2)}$} & 0.10982\rlap{$^{(2)}$} & 0.00897\rlap{$^{(2)}$} & 0.00128\rlap{$^{(2)}$} \\
SiF & 0.03418 & 0.10916 & 0.00743 & 0.00062 \\
N$_2$O & 0.03719\rlap{$^{(2)}$} & 0.10721\rlap{$^{(2)}$} & 0.00740\rlap{$^{(2)}$} & 0.00132\rlap{$^{(4)}$} \\
ClF & 0.02311 & 0.10618 & 0.00304\rlap{$^{(2)}$} & 0.00020\rlap{$^{(5)}$} \\
AlCl & 0.01875\rlap{$^{(2)}$} & 0.10593\rlap{$^{(2)}$} & 0.00166\rlap{$^{(2)}$} & 0.00021\rlap{$^{(2)}$} \\
N$_2$ & 0.01770\rlap{$^{(2)}$} & 0.10487\rlap{$^{(2)}$} & 0.00343\rlap{$^{(2)}$} & 0.00104\rlap{$^{(2)}$} \\
BF & 0.02547\rlap{$^{(2)}$} & 0.10460\rlap{$^{(2)}$} & 0.00281\rlap{$^{(2)}$} & 0.00041\rlap{$^{(2)}$} \\
CN & 0.26336 & 0.10349\rlap{$^{(2)}$} & 0.01677 & 0.00204\rlap{$^{(2)}$} \\
CH & 0.01394 & 0.10343 & 0.00432 & 0.00033 \\
HOF & 0.01755 & 0.10067 & 0.00216 & 0.00012\rlap{$^{(2)}$} \\
C$_2$H & 0.13944\rlap{$^{(2)}$} & 0.10043\rlap{$^{(2)}$} & 0.00388\rlap{$^{(2)}$} & 0.00214 \\
\bottomrule
\end{tabular}
\hfill
\begin{tabular}[t]{@{}l*{4}{r@{\hspace{17pt}}}@{}}
\toprule
Species & {$\max|T_1|$} & {$\max|T_2|$} & {$\max|T_3|$} & {$\max|T_4|$} \\
\midrule
HCN & 0.01694\rlap{$^{(2)}$} & 0.09832\rlap{$^{(2)}$} & 0.00250\rlap{$^{(2)}$} & 0.00070\rlap{$^{(2)}$} \\
SO & 0.03408 & 0.09726 & 0.01055 & 0.00097\rlap{$^{(2)}$} \\
AlH & 0.01348 & 0.09585\rlap{$^{(2)}$} & 0.00207\rlap{$^{(2)}$} & 0.00015\rlap{$^{(4)}$} \\
C$_2$H$_2$ & 0.01791\rlap{$^{(2)}$} & 0.09580\rlap{$^{(2)}$} & 0.00247\rlap{$^{(2)}$} & 0.00075\rlap{$^{(2)}$} \\
SiH & 0.02588 & 0.08570 & 0.00624 & 0.00058 \\
ClO & 0.18296 & 0.07871 & 0.00691 & 0.00047 \\
CO & 0.02748\rlap{$^{(2)}$} & 0.07663\rlap{$^{(2)}$} & 0.00760\rlap{$^{(2)}$} & 0.00115\rlap{$^{(2)}$} \\
SiO & 0.05296\rlap{$^{(2)}$} & 0.07549\rlap{$^{(2)}$} & 0.01176\rlap{$^{(2)}$} & 0.00281\rlap{$^{(2)}$} \\
H$_2$ & 0.00670 & 0.05642 & {---} & {---} \\
OH & 0.00967 & 0.05000 & 0.00441 & 0.00035 \\
H$_2$O & 0.00700 & 0.04976 & 0.00175 & 0.00008\rlap{$^{(2)}$} \\
H$_2$S & 0.01120 & 0.04860 & 0.00112 & 0.00007\rlap{$^{(2)}$} \\
PH$_3$ & 0.01672\rlap{$^{(2)}$} & 0.04741\rlap{$^{(2)}$} & 0.00143\rlap{$^{(2)}$} & 0.00005\rlap{$^{(2)}$} \\
SH & 0.01705 & 0.04679 & 0.00294 & 0.00021 \\
NH & 0.01353 & 0.04633 & 0.00250\rlap{$^{(2)}$} & 0.00024 \\
HCl & 0.00760 & 0.04580 & 0.00154\rlap{$^{(2)}$} & 0.00012\rlap{$^{(2)}$} \\
HF & 0.00772\rlap{$^{(2)}$} & 0.04231 & 0.00229\rlap{$^{(2)}$} & 0.00021\rlap{$^{(4)}$} \\
BH$_3$ & 0.00606 & 0.04097 & 0.00080\rlap{$^{(2)}$} & 0.00006\rlap{$^{(2)}$} \\
CH$_3$ & 0.00961 & 0.03908 & 0.00169\rlap{$^{(2)}$} & 0.00015\rlap{$^{(2)}$} \\
CH$_4$ & 0.00549 & 0.03898 & 0.00070 & 0.00004\rlap{$^{(4)}$} \\
NH$_3$ & 0.00509 & 0.03814 & 0.00079\rlap{$^{(2)}$} & 0.00005\rlap{$^{(2)}$} \\
$^3$CH$_2$ & 0.01805 & 0.03508 & 0.00202 & 0.00024 \\
SiH$_4$ & 0.01463 & 0.03330 & 0.00114 & 0.00005\rlap{$^{(2)}$} \\
AlH$_3$ & 0.01048 & 0.03263 & 0.00101\rlap{$^{(2)}$} & 0.00007\rlap{$^{(2)}$} \\
NH$_2$ & 0.00930 & 0.03111 & 0.00173 & 0.00016 \\
\bottomrule
\end{tabular}
\end{table*}

Tables~\ref{tab:seven_triples} and \ref{tab:seven_quads} present the post-CCSD(T) corrections for seven cases with strong static correlation, namely singlet BN, triplet \ce{B2}, singlet dicarbon, diphosphorus, dinitrogen oxide, ozone, and the \ce{NO2} radical. (By way of background information, Table~\ref{tab:maxamp} presents the largest singles, doubles, triples, and quadruples amplitudes for each species considered in this paper.) We also added to the table the two radicals with the most severe spin contamination, namely CN and CCH.

The reference values are twofold. One set uses the Riemann-3 extrapolation of Lesiuk and co-workers\cite{Lesiuk2018} on sextuple- and septuple-zeta basis sets. The other uses the same extrapolation but on quintuple- and sextuple-zeta basis sets. As a sanity check, we additionally applied the $L^{-3}$ simple extrapolation\cite{Halkier1998a} on sextuple- and septuple-zeta. It can be seen in Tables~\ref{tab:seven_triples} and \ref{tab:seven_quads} that for these specific cases the three extrapolations by and large result in the same values. That is to say that the largest basis sets are large enough that the specific details of the extrapolation method no longer matter for the complete basis set limit estimate.

As can be seen in the top block of Table~\ref{tab:seven_triples}, the CCSDT-3 $-$ CCSD(T) difference converges relatively quickly and smoothly, with the possible exception of the pathological BN diatomic, but even that one has entered an asymptotic regime by quadruple-zeta and especially quintuple-zeta. The basis set convergence slope is positive --- that is to say, the contribution becomes systematically less negative, or in the cases of \ce{NO2} and \ce{CN}  more positive, as the basis set is increased.

In contrast, we see in the CCSDT $-$ CCSDT-3 block that these contributions converge considerably more slowly. Here the basis set convergence slope is negative:  contributions become more negative, or in the cases of \ce{B2}, \ce{CN}, and \ce{CCH} less positive, as the basis set cardinal number is increased. In fact, for \ce{C2} they change sign, from positive to negative. For the other species they are negative. If one compares the septuple-zeta numbers with the estimated basis set limits, it is very clear that  CCSDT $-$ CCSDT-3 is actually the term that is slowing down the convergence of CCSDT $-$ CCSD(T). 

In an attempt to more closely pinpoint the cause of this behavior, we carried out a further decomposition. Greatly expanding on an older study by Kucharski and Bartlett,\cite{Kucharski1986} Cremer and coworkers\cite{Cremer2001} presented a detailed analysis by orders of perturbation theory of the differences between exact CCSDT and various approximations thereto. (For the purpose of the discussion, CCSD(T) can be seen as a noniterative approximation to the CCSDT-1b $-$ CCSD difference.) 
CCSDT-2 iteratively introduces the effect of disconnected quadruples $T_2^2/2$ on the triples amplitudes $T_3$, which starts out at fifth order in perturbation theory with $E^{(5)}_{TQ_{dd}}$ (i.e., the triples-disconnected quadruples term) while CCSDT-3\cite{CCSDT-3} additionally introduces singles $T_1^n/n!$ and cross-products such as $T_1T_2$ --- their (quite weak) influence begins at sixth order in perturbation theory. This is the highest CCSDT-n level that still has the same $O(N^7)$ CPU time scaling as CCSD(T).

CCSDT-4\cite{CCSDT-4andCCSDT} for the first time introduces a linear $T_3$ term into the triples amplitudes equation, allowing for the first time direct triples-triples interactions. The leading term in perturbation theory is $E^{(5)}_{TT}$. 

The remaining difference with full CCSDT consists of nonlinear terms involving $T_3$, such as $T_2T_3$ , $T_1T_3$,  and $T_1^2T_3/2$, which in perturbation theory will affect the energy starting at orders 6, 7, and 9, respectively.

Using the CCSDT-4 implementation in CFOUR's legacy VCC module, we were able to decompose the CCSDT $-$ CCSDT-3 difference into its constituent components through cc-pV6Z. As seen in Table~\ref{tab:seven_triples}, the nonlinear CCSDT $-$ CCSDT-4 difference is quite substantial, but converges rapidly with the basis set, with cc-pV5Z being converged to better than 0.01 \kcalmol across the board.  To be more specific: the cc-pVQZ, cc-pV5Z, and cc-pV6Z sequence indicates roughly $\propto L^{-4}$ convergence pattern, and if we use $L^{-4}$ extrapolation from cc-pV\{5,6\}Z, we find that cc-pV5Z and cc-pV6Z are just RMS 0.007 and 0.003 \kcalmol, respectively, away from the basis set limit. In view of this, CCSDT-4/cc-pV7Z calculations seem entirely redundant, which is just as well as VCC required the legacy integral and SCF codes which `top out' at $i$ functions.

%We can thus conclude that the CCSDT $-$ CCSDT-3 difference for large basis sets is entirely dominated by the `triples on triples' $\hat{W}\hat{T_3}$ contribution, which corresponds to CCSDT-4 $-$ CCSDT-3. 

We can hence conclude that the somewhat sluggish and erratic basis set convergence of the CCSDT $-$ CCSD(T) component is borne of the superposition of three components with different behaviors: a slowly converging linear triples-on-triples $\hat{W}\hat{T_3}$ with a negative basis set convergence slope, a moderately fast-converging CCSDT-3 $-$ CCSD(T) with a countervailing positive slope, and finally a rapidly convergent nonlinear-triples contribution CCSDT $-$ CCSDT-4 with a negative slope. In the large basis set limit, the slow $\hat{W}\hat{T_3}$ entirely sets the pace.

Why is specifically the said triples-on-triples term so slow to converge? For lower excitation levels, it is fairly well established (see, e.g. Irmler, Gr\"uneis, and coworkers\cite{Irmler2019JCP,Irmler2019PRL,Irmler2024}) that ladder diagrams systematically exhibit slower basis set convergence than ring diagrams. As the Feynman-Brandow diagrams for $E^{(5)}_{TT}$ are clearly ladder diagrams, our findings fit that picture.

Empirically, it has been known for some time that the total `$T_3-(T)$' contribution, CCSDT $-$ CCSD(T), converges more slowly with the basis set than the idealized $\propto L^{-3}$ behavior; Karton et al.\cite{jmlm205} found approximately $\propto L^{-2.5}$ for smaller basis sets, while a decade and a half later, for a small sample and basis sets up to cc-pV6Z,  Karton\cite{Karton2020} found inverse-$L$ exponents in the 2.5--2.75 range depending on the basis set pair. 

The above suggests that, for accurate work, there may be some merit in partitioning the difference into two components, namely. CCSDT-3 $-$ CCSD(T) and CCSDT $-$ CCSDT-3. This introduces another $O(N^7)$ calculation but avoids an additional $O(N^8)$ CCSDT-4 step, and basis set convergence of the nonlinear $T_3$ terms appears to be rapid enough that we can `bury it in the mix'.

%The superposition of the two trends in this third block somewhat masks the basis set convergence behavior. Here too, we still have a fairly significant difference between the largest basis set and the extrapolated limit.

\subsection{A brief intermezzo on CCSDT-3 $-$ CCSDT-2}

Somewhat analogous to the difference between CCSDT and CCSDT-4, the mind wonders what is the effect of the $T_1$, $T_1T_2$, etc. terms in the $T_3$ amplitudes equation that make up the difference between CCSDT-3 and the simpler CCSDT-2. 

First of all, we assessed the magnitude of the CCSDT-2 -- CCSD(T) difference for the entire W4-17 dataset, using the cc-pVTZ basis set. All contributions to TAE were found to be negative except for three: SiF where it is less than 0.01 \kcalmol anyhow, CF where it is a picayune +0.040 \kcalmol, and SiO, which has the only nontrivially positive contribution at 0.196 \kcalmol. For the cc-pVQZ and cc-pV5Z this is somewhat reduced to 0.158 and 0.159 \kcalmol, respectively, but clearly remains positive.

In contrast, CCSDT-3 $-$ CCSDT-2 for all these species is positive across the board. Conspicuously large contributions are seen for radicals with strong UHF spin contamination (e.g., Table~\ref{tab:seven_triples}) such as 1.1 \kcalmol for CN and 0.5 \kcalmol for CCH. But significantly positive contributions are also seen for species with multiple bonds involved O, F, Cl, and S. The combination can cause some species (such as \ce{NO2} and \ce{CN} in Table~\ref{tab:seven_triples}) to have an overall positive CCSDT-3 $-$ CCSD(T), bucking the general trend.

Basis set convergence of this contribution is quite rapid, with the  motion across cc-pVQZ, cc-pV5Z, and cc-pV6Z corresponding to a rough apparent $\propto L^{-5}$ convergence behavior, with a systematically upward `basis set slope'. If we carry out an $L^{-5}$ extrapolation from cc-pV\{5,6\}Z, it is found to bridge only 0.002 \kcalmol RMSD from cc-pV6Z to the CBS limit; for cc-pV5Z to the CBS limit, the corresponding figure is still only a measly 0.005 \kcalmol.   While separate evaluation of CCSDT-3 $-$ CCSDT-2 offers no computational benefit, it does allow for rationalizing the seemingly counterintuitive positive (binding) CCSDT-3 $-$ CCSD(T) contributions seen for some species.

\subsection{Basis set convergence: quadruples}

In contrast, (Q) converges much more smoothly and rapidly. Already the cc-pVTZ basis set is within about 10\% of the basis set limit values. For BN, \ce{B2}, \ce{C2}, and \ce{P2}, we have (Q) values with both sextuple-zeta and septuple-zeta,  And the three differ by well below 0.01 \kcalmol. 
As a result, a Riemann3 pV\{5,6\}Z extrapolation only has to bridge a small amount, and where we have pV7Z available pV\{5,6\}Z (`REF2') and pV\{6,7\}Z (`REF1') differ by just a few thousandths of a \kcalmol. 
Hence, we feel confident in using REF2 as the basis set limit for quadruples.

Of course, the  \emph{converged} (Q) contributions \emph{themselves}  are quite significant, reaching over three \kcalmol for the BN and \ce{C2} diatomics, almost two and a half \kcalmol for \ce{N2O} and \ce{NO2}, and almost five \kcalmol for ozone.

%\textbf{[Note to self: Backfill sextuple-zeta number for \ce{NO2} when available.]}

Now what about still higher-order effects? Stanton and co-workers have already made a clear case\cite{jmlm326} that (Q)$_\Lambda$ is in fact a superior method over (Q), and that in general the parenthetical $\Lambda$ sequence CCSD(T)$_\Lambda$, CCSDT(Q)$_\Lambda$, \ldots converges more smoothly to the complete coupled cluster limit than the parenthetical sequence CCSD(T), CCSDT(Q), etc. Hence, in W5preview1, we actually proposed\cite{jmlm340} using (Q)$_\Lambda$ as the highest level of theory.

Now, since the evaluation of (Q)$_\Lambda$ requires the solution of the left eigenvector --- which about triples or quadruples total CPU time and greatly increases memory requirements --- we sought to do this in a relatively small basis set. As can be seen in Table \ref{tab:seven_quads}, even with the cc-pVDZ basis set, the values obtained are of decent quality. The largest basis set for which we were able to compute them for most of the sample was the truncated cc-pV5Z(g) basis set, which we denoted `REF3'. Contributions here can get fairly large for molecules with significant static correlation, such as ozone (-0.54 \kcalmol) \ce{C2} (-0.6 \kcalmol) and \ce{BN} (-0.9 \kcalmol).

However, even the truncated triple-zeta basis set cc-pVTZ(d), where we removed the f functions, approaches REF3 quite well, and untruncated cc-pVTZ does better still.

It is remarkable, incidentally, that the performance of quadruple-zeta truncated at f functions and that of the untruncated cc-pVQZ  basis set is essentially identical. There is a little bit more daylight between truncated triple-zeta and complete triple-zeta, but not a whole lot. Between quintuple-zeta truncated at f and g functions, we are talking about 0.003 kilocalorie per mole or less. This hints at radial flexibility in the basis set being much more important than angular flexibility for these high-order corrections.
Turning now to CCSDTQ $-$ CCSDT(Q)$_\Lambda$, here we see essentially the same tendencies.

\begin{table}
  \caption{RMSD and MSD (\kcalmol) of basis-set results for the valence correlation
    contributions to the TAE. Rows are sorted by descending RMSD[(T)]}
  \label{tab:statsccsdt}
  \footnotesize
\setlength{\tabcolsep}{1.5pt}
\begin{tabular}{@{}l S[table-format=2.2] S[table-format=-2.2] S[table-format=2.2] S[table-format=-2.2] S[table-format=1.3] S[table-format=-1.3]@{}}
\toprule
 & \multicolumn{2}{c}{$[\textrm{E2}]$} & \multicolumn{2}{c}{$[\textrm{CCSD}]_\textrm{corr}$} & \multicolumn{2}{c@{}}{$[(\textrm{T})]$} \\
\cmidrule(lr){2-3}\cmidrule(lr){4-5}\cmidrule(l){6-7}
Basis & {RMSD} & {MSD} & {RMSD} & {MSD} & {RMSD} & {MSD} \\
\midrule
VDZ & 15.81 & -14.56 & 12.07 & -11.07 & 2.443 & -1.992 \\
VTZ(d) & 11.91 & -10.80 & 10.54 & -9.52 & 0.971 & -0.779 \\
VQZ(d) & 10.28 & -9.33 & 9.13 & -8.23 & 0.740 & -0.576 \\
V5Z(d) & 9.65 & -8.76 & 8.60 & -7.74 & 0.647 & -0.494 \\
V7Z(d) & 9.02 & -8.19 & 8.02 & -7.21 & 0.591 & -0.443 \\
VTZ & 7.24 & -6.52 & 5.71 & -5.13 & 0.724 & -0.599 \\
haVTZ & 5.92 & -5.33 & 4.70 & -4.21 & 0.575 & -0.471 \\
pwCVTZ\rlap{$^{a}$} & 7.08 & -6.39 & 5.65 & -5.09 & 0.579 & -0.484 \\
VQZ(f) & 5.25 & -4.65 & 4.23 & -3.70 & 0.418 & -0.338 \\
VQZ & 3.40 & -3.02 & 2.45 & -2.15 & 0.329 & -0.267 \\
V5Z(f) & 4.49 & -3.95 & 3.67 & -3.18 & 0.309 & -0.244 \\
haVQZ & 2.72 & -2.42 & 1.90 & -1.67 & 0.256 & -0.207 \\
V7Z(f) & 3.87 & -3.37 & 3.19 & -2.74 & 0.252 & -0.194 \\
pwCVQZ\rlap{$^{a}$} & 3.31 & -2.95 & 2.43 & -2.15 & 0.266 & -0.219 \\
V5Z(g) & 2.51 & -2.20 & 1.83 & -1.58 & 0.200 & -0.160 \\
V5Z & 1.85 & -1.64 & 1.28 & -1.12 & 0.165 & -0.133 \\
haV5Z & 1.51 & -1.35 & 1.02 & -0.90 & 0.132 & -0.107 \\
V7Z(g) & 1.91 & -1.64 & 1.42 & -1.21 & 0.138 & -0.108 \\
V6Z$^{b}$ & 1.13 & -1.00 & 0.78 & -0.68 & 0.100 & -0.081 \\
V7Z$^{c}$ & 0.76 & -0.67 & 0.55 & -0.48 & 0.076 & -0.063 \\
\bottomrule
\end{tabular}

\smallskip
\parbox{\columnwidth}{\footnotesize
Deviations relative to REF2 (Riemann3 cc-pV\{5,6\}Z extrapolation), $n=49$ unless noted; MSD${}<0$: basis value too small. V$n$Z = cc-pV$n$Z; V$n$Z($l$): truncated at angular momentum $l$ (hydrogen one $l$ lower).
$^{a}$\,Core--valence basis set; valence electrons correlated only, as for all other entries.
$^{b}$\,Not independent of REF2.
$^{c}$\,$n=42$.}
\end{table}

\subsection{Statistics for the whole test set}

We shall now move on from looking at some individual cases to the statistics over the whole dataset. First of all, for perspective, results for MP2, CCSD, and (T) can be found in Table~\ref{tab:statsccsdt}. It is quite clear there that for MP2 and CCSD, angular flexibility going from double-zeta to triple-zeta, including f functions, to quadruple-zeta, including g functions, to quintuple-zeta, including h functions, and so on and so forth, is the main driver of basis set convergence, with added radial flexibility, as measured by the progression, for instance, from quadruple-zeta to quintuple-zeta truncated at f, to septuple-zeta truncated at f, and similar sequences, being much less important.

For the (T) contribution, however, as already pointed out in a book chapter by Martin\cite{jmlm301}, radial flexibility is a lot more impactful, and for example, the RMSD from cc-pVDZ to triple-zeta truncated at d is cut by more than half, and ultimately from septuple-zeta truncated at d, it is down to 0.6 kilocalorie per mole versus 0.7 for the cc-pVTZ basis set. Likewise, cc-pV5Z(f,d) is actually superior for triples to untruncated cc-pVQZ.

\begin{table*}[!t]
  \caption{RMSD and MSD (\kcalmol) of basis-set results for post-CCSD(T) contributions
    to the TAE, relative to the reference limits. Rows are sorted by descending RMSD for CCSDT$-$CCSDT-3}
  \label{tab:statspost}
  \footnotesize
\setlength{\tabcolsep}{3pt}
\begin{tabular*}{\textwidth}{@{\extracolsep{\fill}}l S[table-format=1.3] S[table-format=-1.3] S[table-format=1.3] S[table-format=-1.3] S[table-format=1.3] S[table-format=-1.3] S[table-format=1.3] S[table-format=-1.3] S[table-format=1.4] S[table-format=-1.4] S[table-format=1.4] S[table-format=-1.4]@{}}
\toprule
 & \multicolumn{2}{c}{\shortstack{$[\textrm{CCSDT-3}]$\\$-[\textrm{CCSD(T)}]$}} & \multicolumn{2}{c}{\shortstack{$[\textrm{CCSDT}]$\\$-[\textrm{CCSDT-3}]$}} & \multicolumn{2}{c}{\shortstack{$[\textrm{CCSDT}]$\\$-[\textrm{CCSD(T)}]$}} & \multicolumn{2}{c}{$[(\textrm{Q})]$} & \multicolumn{2}{c}{\shortstack{$[(\textrm{Q})_\Lambda]$\\$-[(\textrm{Q})]$}} & \multicolumn{2}{c@{}}{\shortstack{$[\textrm{CCSDTQ}]$\\$-[\textrm{CCSDT(Q)}_\Lambda]$}} \\
\cmidrule(lr){2-3}\cmidrule(lr){4-5}\cmidrule(lr){6-7}\cmidrule(lr){8-9}\cmidrule(lr){10-11}\cmidrule(l){12-13}
 & \multicolumn{6}{c}{vs.\ REF2} & \multicolumn{2}{c}{vs.\ REF2$^{d}$} & \multicolumn{2}{c}{vs.\ REF3} & \multicolumn{2}{c@{}}{vs.\ REF4$^{e}$} \\
Basis & {RMSD} & {MSD} & {RMSD} & {MSD} & {RMSD} & {MSD} & {RMSD} & {MSD} & {RMSD} & {MSD} & {RMSD} & {MSD} \\
\midrule
VDZ & 0.125 & -0.012 & 0.575 & 0.425 & 0.526 & 0.413 & 0.267 & -0.158 & 0.0358 & 0.0108 & 0.0137 & 0.0000 \\
VTZ(d) & 0.113 & -0.064 & 0.344 & 0.265 & 0.252 & 0.202 & 0.197 & -0.142 & 0.0132 & 0.0032 & 0.0093 & 0.0002 \\
VQZ(d) & 0.106 & -0.062 & 0.288 & 0.224 & 0.204 & 0.162 & 0.160 & -0.114 & 0.0105 & 0.0017 & 0.0073 & -0.0015 \\
V5Z(d) & 0.101 & -0.058 & 0.270 & 0.209 & 0.191 & 0.151 & 0.147 & -0.105 & 0.0095 & 0.0012 & 0.0073 & -0.0018 \\
V7Z(d) & 0.098 & -0.056 & 0.262 & 0.202 & 0.186 & 0.147 & 0.140 & -0.100 & 0.0089 & 0.0008 & 0.0072 & -0.0020 \\
VTZ & 0.086 & -0.057 & 0.257 & 0.191 & 0.177 & 0.134 & 0.119 & -0.085 & 0.0092 & 0.0014 & 0.0063 & 0.0041 \\
haVTZ & 0.116 & -0.076 & 0.216 & 0.158 & 0.109 & 0.082 & 0.079 & -0.059 & {---} & {---} & {---} & {---} \\
pwCVTZ\rlap{$^{a}$} & 0.059 & -0.039 & 0.203 & 0.151 & 0.150 & 0.112 & 0.103 & -0.074 & 0.0072 & 0.0002 & {---} & {---} \\
VQZ(f) & 0.067 & -0.047 & 0.159 & 0.117 & 0.095 & 0.070 & 0.064 & -0.046 & 0.0033 & -0.0000 & 0.0014 & 0.0008 \\
VQZ & 0.050 & -0.036 & 0.136 & 0.103 & 0.088 & 0.066 & 0.049 & -0.035 & 0.0021 & 0.0001 & 0.0019 & -0.0009 \\
V5Z(f) & 0.058 & -0.040 & 0.125 & 0.090 & 0.071 & 0.050 & 0.047 & -0.033 & 0.0018 & -0.0004 & {---} & {---} \\
haVQZ & 0.057 & -0.039 & 0.119 & 0.090 & 0.065 & 0.050 & 0.032 & -0.024 & {---} & {---} & {---} & {---} \\
V7Z(f) & 0.052 & -0.035 & 0.108 & 0.075 & 0.059 & 0.040 & 0.039 & -0.027 & {---} & {---} & {---} & {---} \\
pwCVQZ\rlap{$^{a}$} & 0.035 & -0.026 & 0.107 & 0.080 & 0.073 & 0.054 & 0.043 & -0.031 & {---} & {---} & {---} & {---} \\
V5Z(g) & 0.036 & -0.026 & 0.090 & 0.067 & 0.055 & 0.041 & 0.028 & -0.020 & {---} & {---} & {---} & {---} \\
V5Z & 0.029 & -0.021 & 0.077 & 0.058 & 0.049 & 0.037 & 0.024 & -0.017 & {---} & {---} & {---} & {---} \\
haV5Z & 0.029 & -0.021 & 0.067 & 0.051 & 0.039 & 0.030 & 0.017 & -0.012 & {---} & {---} & {---} & {---} \\
V7Z(g) & 0.028 & -0.020 & 0.065 & 0.046 & 0.038 & 0.026 & 0.019 & -0.014 & {---} & {---} & {---} & {---} \\
V6Z\rlap{$^{b}$} & 0.017 & -0.013 & 0.047 & 0.036 & 0.030 & 0.023 & 0.015 & -0.010 & {---} & {---} & {---} & {---} \\
V7Z\rlap{$^{c}$} & 0.012 & -0.009 & 0.032 & 0.025 & 0.021 & 0.016 & 0.010 & -0.008 & {---} & {---} & {---} & {---} \\
\midrule
\multicolumn{13}{@{}l}{\textit{Two-point extrapolations}} \\
V\{D,T\}Z, $\alpha=3$ & 0.096 & -0.076 & 0.127 & 0.093 & 0.047 & 0.017 & 0.080 & -0.054 & {---} & {---} & {---} & {---} \\
$1.10\times$VTZ & {} & {} & {} & {} & {} & {} & 0.040 & -0.013 & {} & {} & {} & {} \\
V\{T,Q\}Z, $\alpha=3$ & 0.028 & -0.021 & 0.050 & 0.038 & 0.025 & 0.017 & 0.008 & 0.002 & {---} & {---} & {---} & {---} \\
V\{T,Q\}Z, $L_\textrm{eff}$\rlap{$^{f}$} & 0.024 & -0.006 & 0.018 & 0.001 & 0.011 & -0.002 & 0.008 & -0.003 & {---} & {---} & {---} & {---} \\
V\{Q,5\}Z, $\alpha=3$ & 0.007 & -0.005 & 0.016 & 0.012 & 0.010 & 0.007 & 0.004 & 0.002 & {---} & {---} & {---} & {---} \\
V\{Q,5\}Z, $L_\textrm{eff}$\rlap{$^{f}$} & 0.005 & 0.000 & 0.009 & -0.000 & 0.008 & 0.000 & 0.004 & -0.002 & {---} & {---} & {---} & {---} \\
\bottomrule
\end{tabular*}

\smallskip
\parbox{\textwidth}{\footnotesize
Valence electrons correlated; W4-17 subset of 49 species (H$_2$, HOO, HOOH, F$_2$O, HOCl, HSS excluded). Deviation = basis $-$ reference; MSD${}>0$: basis value too positive. V$n$Z = cc-pV$n$Z; V$n$Z($l$): truncated at angular momentum $l$ (hydrogen one $l$ lower). REF2: Riemann3 cc-pV\{5,6\}Z extrapolation; REF3 = V5Z(g); REF4 = V5Z(f). ---: no data, or the basis is itself the reference.
$^{a}$\,Core--valence basis set, valence electrons correlated only.
$^{b}$\,Not independent of REF2.
$^{c}$\,$n=42$; $n=34$ for $[(\textrm{Q})]$.
$^{d}$\,$n=49$.
$^{e}$\,$n=45$.
$^{f}$\,Two-point coefficient $\lambda$ from the effective cardinal numbers of Table~X (fitted to these data, hence somewhat optimistic; compare the $\alpha=3$ rows).}
\end{table*}

Let us now turn to the post-CCSD(T) statistics in Table~\ref{tab:statspost}. For CCSDT-3 $-$ CCSD(T), we see a relatively rapid convergence by angular momenta. We also see that adding radial flexibility helps a little, but not as much as does adding angular momenta.

The behavior of CCSDT $-$ CCSDT-3 is quite different. Here, the RMSD for double-zeta is almost fivefold the previous contribution. But in addition, increasing the radial flexibility is quite impactful, with, for example, the septuple-zeta truncated at d functions having an RMS error that is less than half that of an ordinary cc-pVDZ basis set, and the quadruple-zeta truncated at t being nearly as good.

Similarly, nearly all of the difference between cc-pVTZ and cc-pVQZ is captured by the quadruple-zeta basis set truncated at f functions. And the quintuple-zeta truncated at f functions actually outperforms the quadruple-zeta, which includes g functions. Similarly, a quintuple-zeta truncated at g is quite close in performance to the full quintuple-zeta.

It is clear that radial flexibility plays a greater role for this important term. And once you add up the two contributions in CCSDT $-$ CCSD(T), that behavior is borne out too.

For the parenthetical quadruples, (Q), once again we see that, important as angular momentum flexibility seems to be at first sight, detailed analysis reveals that including the radial flexibility is almost more important. To wit, cc-pVDZ at 0.266 \kcalmol RMS drops to 0.138 for septuple-zeta truncated at d functions, whereas cc-pVTZ is only slightly below that at 0.115.

Quadruple-zeta truncated at f functions clocks in at 0.06 \kcalmol, compared to 0.05 for the complete cc-pVQZ basis set. Quintuple-zeta truncated at f: 0.046, truncated at g: 0.028, complete quintuple-zeta basis set: 0.023. It would only be a mild exaggeration to say that the top angular momentum can safely be removed.

Finally, the complete septuple-zeta basis set is just 0.01 \kcalmol from the extrapolated limit.

For CCSDT(Q)$_\Lambda$ $-$ CCSDT(Q), going from a double-zeta to a triple-zeta truncated at d functions basis set reduces the already small RMSD of 0.036 \kcalmol by two-thirds. And the remaining value of 0.013 is only slightly less than the cc-pVDZ at 0.01 \kcalmol. Walking further up, differences become essentially meaningless.
For CCSDTQ $-$ CCSDT(Q)$_\Lambda$, one can likewise safely truncate at TZ, truncated at d functions, but even the cc-pVDZ there would have been more than adequate, at an RMSD of 0.014 \kcalmol.

%\subsection{Splice this in somewhere appropriate}

\begin{table*}[!t]
  \caption{RMS (\kcalmol) of the TAE increments from successive shell additions along
    cc-pVTZ $\to$ cc-pVQZ, their relative shares, and the effect of the tight functions
    of cc-pwCVTZ; valence electrons correlated.}
  \label{tab:increments}
  
\footnotesize
\setlength{\tabcolsep}{3pt}
\begin{tabular}{@{}l S[table-format=1.4] S[table-format=1.4] S[table-format=1.4] S[table-format=1.4] c c c c S[table-format=1.4]@{}}
\toprule
 & \multicolumn{4}{c}{RMS of step (kcal/mol)} & \multicolumn{4}{c}{share} & {RMS} \\
\cmidrule(lr){2-5}\cmidrule(lr){6-9}
 & {+sp} & {+d} & {+f} & {+g} & sp & d & f & g & {+t(spdf)} \\
\midrule
$[\textrm{E2}]$ & 0.418 & 0.575 & 1.130 & 1.859 & 10\% & 14\% & 28\% & 47\% & 0.220 \\
$[\textrm{CCSD}]-[\textrm{MP2}]$ & 0.157 & 0.237 & 0.222 & 0.152 & 20\% & 31\% & 29\% & 20\% & 0.134 \\
$[\textrm{CCSD}]_\textrm{corr}$ & 0.340 & 0.360 & 0.934 & 1.797 & 10\% & 10\% & 27\% & 52\% & 0.145 \\
$[(\textrm{T})]$ & 0.111 & 0.098 & 0.108 & 0.092 & 27\% & 24\% & 26\% & 23\% & 0.152 \\
$[\textrm{CCSDT-3}]-[\textrm{CCSD(T)}]$ & 0.009 & 0.009 & 0.012 & 0.018 & 19\% & 18\% & 26\% & 37\% & 0.030 \\
$[\textrm{CCSDT-4}]-[\textrm{CCSDT-3}]$ & 0.021 & 0.029 & 0.027 & 0.016 & 23\% & 31\% & 29\% & 17\% & 0.046 \\
$[\textrm{CCSDT}]-[\textrm{CCSDT-4}]$ & 0.012 & 0.010 & 0.009 & 0.010 & 28\% & 25\% & 23\% & 24\% & 0.012 \\
$[\textrm{CCSDT}]-[\textrm{CCSDT-3}]$ & 0.029 & 0.038 & 0.035 & 0.024 & 23\% & 30\% & 28\% & 19\% & 0.055 \\
$[\textrm{CCSDT}]-[\textrm{CCSD(T)}]$ & 0.023 & 0.038 & 0.026 & 0.012 & 24\% & 38\% & 26\% & 12\% & 0.029 \\
$[(\textrm{Q})]$ & 0.021 & 0.018 & 0.018 & 0.016 & 29\% & 24\% & 25\% & 22\% & 0.016 \\
$[(\textrm{Q})_\Lambda]-[(\textrm{Q})]$ & 0.0040 & 0.0024 & 0.0020 & 0.0015 & 40\% & 24\% & 20\% & 15\% & 0.0033 \\
$[\textrm{CCSDTQ}]-[\textrm{CCSDT(Q)}_\Lambda]$\rlap{$^{a}$} & 0.0029 & 0.0014 & 0.0020 & 0.0029 & 31\% & 15\% & 23\% & 31\% & {---} \\
\bottomrule
\end{tabular}

\smallskip
\parbox{\textwidth}{\footnotesize
SD = CCSD, T-3 = CCSDT-3, T = CCSDT, SD(T) = CCSD(T), TQ = CCSDTQ, T(Q)$_\Lambda$ = CCSDT(Q)$_\Lambda$. RMS over 49 species of the TAE increments along cc-pVTZ $\to$ VTZ+sp $\to$ VTZ+spd $\to$ VQZ(f) $\to$ cc-pVQZ (columns +sp, +d, +f, +g); share = step RMS divided by the sum of the four. +t(spdf): cc-pwCVTZ $-$ cc-pVTZ (valence correlation only). 
$^{a}$\,47 species (no CCSDTQ/cc-pVQZ for NO$_2$ and O$_3$); +t(spdf) not available.}
\end{table*}

Another perspective is offered in Table~\ref{tab:increments}, where we 
%In an attempt to narrow down the effect of adding additional radial flexibility, we 
%split up the cc-pwCVTZ and cc-pwCVQZ basis sets in angular momentum layers. 
gradually expanded the cc-pVTZ basis set to cc-pVQZ, taking one angular momentum at a time from the latter. 
We found that for MP2 and CCSD, the main improvement comes from the top angular momentum. In contrast, for CCSDT - CCSDT-3 and higher, the top angular momentum tapers off from about equal importance with the lower angular momentum additions to considerably smaller.

\subsection{Basis set extrapolation}

There are a number of basis set extrapolation formulas in common use. First of all, we must recall that any two-point extrapolation formula can be brought into the linear Schwenke form,\cite{Schwenke2005} 
$E_{\mathrm{CBS}}\approx E(L)+A_L[E(L)-E(L-1)]$. 
And hence all two-point formulas can be made equivalent\cite{jmlm280,IronBloke} with appropriate parametrization. That includes, among others, the variable inverse exponent formula, $L^{-\alpha}$, introduced by Martin and Taylor,\cite{jmlm096} the shifted origin formula, $(L+a)^{-3}$, due to George A. Petersson (1942-2026),\cite{ranasinghe2013ccsd} and others.

Related to the Petersson formula is the Varandas concept\cite{Varandas2014} of extrapolating $X^{-3}$, where $X$ is a cardinal number optimized to the basis set, not necessarily equal to its nominal cardinal number. The determination of these cardinal numbers is an underdetermined problem. In the present work, we adopted the expedient of using a constant shift $a$ for a given energy difference, and then arbitrarily assigning the effective cardinal numbers 4 and 5 to cc-pVQZ and cc-pV5Z, respectively. This approach permits us to assign cardinal numbers also to truncated basis sets or diffuse function augmented ones.

\begin{table*}[!t]
  \caption{Effective cardinal numbers $L_\textrm{eff}$ and RMSD (\kcalmol) for partial
    basis-set expansions between cc-pVTZ and cc-pVQZ; valence electrons correlated.}
  \label{tab:partial}
  \footnotesize
\setlength{\tabcolsep}{3pt}
\begin{tabular*}{\textwidth}{@{\extracolsep{\fill}}l S[table-format=1.4] S[table-format=1.4] S[table-format=1.4] S[table-format=1.4] S[table-format=1.4] S[table-format=1.4] S[table-format=1.4] S[table-format=1.4] c@{}}
\toprule
 & {VTZ} & {VTZ+sp} & {VTZ+spd} & {VTZ+t(sp)} & {VTZ+t(df)} & {pwCVTZ} & {VQZ(f)} & {VQZ} & {$n$} \\
$N_{\rm contr}$ row1/row2 & \multicolumn{1}{c}{30/34} & \multicolumn{1}{c}{34/38} & \multicolumn{1}{c}{39/43} & \multicolumn{1}{c}{38/42} & \multicolumn{1}{c}{35/51} & \multicolumn{1}{c}{43/59} & \multicolumn{1}{c}{46/50} & \multicolumn{1}{c}{55/59} & \\
\midrule
\multicolumn{10}{@{}l}{\textit{Effective cardinal number $L_\textrm{eff}$}} \\
$[\textrm{E2}]$ & 3.02 & 3.07 & 3.16 & 3.02 & 3.04 & 3.05 & 3.39 & 4.00 & \\
$[\textrm{CCSD}]-[\textrm{MP2}]$ & 3.14 & 3.25 & 3.52 & 3.16 & 3.24 & 3.27 & 3.85 & 4.00 & \\
$[\textrm{CCSD}]_\textrm{corr}$ & 3.03 & 3.07 & 3.12 & 3.03 & 3.03 & 3.04 & 3.32 & 4.00 & \\
$[(\textrm{T})]$ & 3.15 & 3.30 & 3.46 & 3.26 & 3.23 & 3.37 & 3.71 & 4.00 & \\
$[\textrm{CCSDT-3}]-[\textrm{CCSD(T)}]$ & 3.19 & 3.30 & 3.32 & 3.48 & 3.41 & 3.80 & 3.53 & 4.00 & \\
$[\textrm{CCSDT}]-[\textrm{CCSDT-3}]$ & 3.09 & 3.24 & 3.48 & 3.24 & 3.22 & 3.41 & 3.76 & 4.00 & \\
$[\textrm{CCSDT}]-[\textrm{CCSD(T)}]$ & 3.03 & 3.19 & 3.54 & 3.13 & 3.13 & 3.24 & 3.88 & 4.00 & \\
$[(\textrm{Q})]$ & 3.10 & 3.22 & 3.39 & 3.15 & 3.19 & 3.24 & 3.64 & 4.00 & \\
\midrule
\multicolumn{10}{@{}l}{\textit{RMSD (kcal/mol) relative to REF1/REF2}} \\
$[\textrm{E2}]$ & 7.23 & 6.87 & 6.33 & 7.22 & 7.08 & 7.06 & 5.23 & 3.38 & 49 \\
$[\textrm{CCSD}]-[\textrm{MP2}]$ & 1.66 & 1.57 & 1.34 & 1.64 & 1.56 & 1.55 & 1.13 & 1.03 & 49 \\
$[\textrm{CCSD}]_\textrm{corr}$ & 5.64 & 5.36 & 5.06 & 5.63 & 5.59 & 5.58 & 4.16 & 2.38 & 49 \\
$[(\textrm{T})]$ & 0.710 & 0.604 & 0.510 & 0.630 & 0.648 & 0.565 & 0.404 & 0.314 & 49 \\
$[\textrm{CCSDT-3}]-[\textrm{CCSD(T)}]$ & 0.088 & 0.081 & 0.079 & 0.073 & 0.076 & 0.060 & 0.069 & 0.052 & 49 \\
$[\textrm{CCSDT}]-[\textrm{CCSDT-3}]$ & 0.258 & 0.231 & 0.195 & 0.231 & 0.234 & 0.205 & 0.161 & 0.138 & 49 \\
$[\textrm{CCSDT}]-[\textrm{CCSD(T)}]$ & 0.178 & 0.157 & 0.121 & 0.164 & 0.164 & 0.151 & 0.096 & 0.089 & 49 \\
$[(\textrm{Q})]$ & 0.117 & 0.097 & 0.080 & 0.110 & 0.108 & 0.101 & 0.062 & 0.047 & 49 \\
$[(\textrm{Q})_\Lambda]-[(\textrm{Q})]$\rlap{$^{a}$} & 0.0092 & 0.0071 & 0.0051 & 0.0084 & 0.0079 & 0.0072 & 0.0033 & 0.0021 & 49 \\
%REF4=cc-pV5Z(f)
%$[\textrm{TQ}]-[\textrm{T(Q)}_\Lambda]$\rlap{$^{b}$} & 0.0063 & 0.0041 & 0.0031 & {---} & {---} & {---} & 0.0014 & 0.0019 & 45 \\
%REF4=cc-pVQZ
$[\textrm{CCSDTQ}]-[\textrm{CCSDT(Q)}_\Lambda]$\rlap{$^{b}$} & 0.0084 & 0.0058 & 0.0047 & {---} & {---} & {---} & 0.0029 & {REF4} & 47 \\
\bottomrule
\end{tabular*}

\smallskip
\parbox{\textwidth}{\footnotesize
SD = CCSD, T-3 = CCSDT-3, T = CCSDT, SD(T) = CCSD(T). Second header row: basis functions per first-row/second-row atom. VTZ = cc-pVTZ; VTZ+sp: s,p shells from cc-pVQZ; VTZ+spd: s,p,d shells from cc-pVQZ; VTZ+t(sp), VTZ+t(df): cc-pVTZ plus the tight s,p or tight d,f functions of cc-pwCVTZ (first row: 2s2p and 1d; second row: 2s2p and 2d1f); VQZ(f) = cc-pVQZ without g functions. Valence electrons correlated throughout. Reference: REF1 (Riemann3 cc-pV\{6,7\}Z) where available, otherwise REF2 (Riemann3 cc-pV\{5,6\}Z). $L_\textrm{eff}$ projected onto the fit of Table~\ref{tab:leff}. $^{a}$\,Relative to REF3 = cc-pV5Z(g).
$^{b}$\,Relative to REF4 = 
%either
%cc-pV5Z(f)
%or
cc-pVQZ.
}
\end{table*}

%begin SVD bit
For each component, we assume that the basis-set deficit of species s in basis set b factorizes as $\delta_sb = A_s w_b$: all species share a common convergence profile across basis sets, $w_b$, and differ only by a species prefactor $A_s$. The prefactors and basis-set weights were obtained from the leading term of a singular value decomposition (SVD) of the matrix of deficits relative to REF1, which is the best rank-one approximation in the least-squares sense. Writing $w_b \propto [L_eff(b) + a]^{-3}$ then defines an effective cardinal number for each basis set, with the scale and the shift a fixed by requiring $L_{\rm eff}$ = 4 for cc-pVQZ and 5 for cc-pV5Z.

%end SVD bit

\begin{table*}[!t]
  \caption{Effective cardinal numbers $L_\textrm{eff}$ and shifts $a$ per TAE component.}
  \label{tab:leff}
  \footnotesize
\setlength{\tabcolsep}{4pt}
\begin{tabular*}{\textwidth}{@{\extracolsep{\fill}}l S[table-format=1.2] S[table-format=1.2] S[table-format=1.2] S[table-format=1.2] S[table-format=1.2] S[table-format=1.2] S[table-format=1.2] S[table-format=1.2]@{}}
\toprule
 & {$[\textrm{E2}]$} & {\shortstack{$[\textrm{CCSD}]$\\$-[\textrm{MP2}]$}} & {$[\textrm{CCSD}]_\textrm{corr}$} & {$[(\textrm{T})]$} & {\shortstack{$[\textrm{CCSDT-3}]$\\$-[\textrm{CCSD(T)}]$}} & {\shortstack{$[\textrm{CCSDT}]$\\$-[\textrm{CCSDT-3}]$}} & {\shortstack{$[\textrm{CCSDT}]$\\$-[\textrm{CCSD(T)}]$}} & {$[(\textrm{Q})]$} \\
\midrule
$n$ & \multicolumn{1}{c}{42} & \multicolumn{1}{c}{42} & \multicolumn{1}{c}{42} & \multicolumn{1}{c}{42} & \multicolumn{1}{c}{42} & \multicolumn{1}{c}{42} & \multicolumn{1}{c}{42} & \multicolumn{1}{c}{33} \\
shift $a$ & \multicolumn{1}{c}{$\mathbf{+0.45}$} & \multicolumn{1}{c}{$\mathbf{+1.87}$} & \multicolumn{1}{c}{$\mathbf{-0.04}$} & \multicolumn{1}{c}{$\mathbf{-0.40}$} & \multicolumn{1}{c}{$\mathbf{+1.10}$} & \multicolumn{1}{c}{$\mathbf{+0.84}$} & \multicolumn{1}{c}{$\mathbf{+0.69}$} & \multicolumn{1}{c}{$\mathbf{-0.66}$} \\
\midrule
VDZ\rlap{$^{a}$} & 2.26 & 1.80 & 2.43 & 2.23 & 3.47 & 2.17 & 1.91 & 2.45 \\
VTZ(d)\rlap{$^{a}$} & 2.50 & 3.19 & 2.49 & 2.89 & 2.89 & 2.74 & 2.65 & 2.66 \\
VQZ(d)\rlap{$^{a}$} & 2.65 & 3.51 & 2.60 & 3.13 & 2.95 & 2.97 & 2.93 & 2.79 \\
V5Z(d)\rlap{$^{a}$} & 2.71 & 3.65 & 2.66 & 3.26 & 3.00 & 3.07 & 3.03 & 2.84 \\
VTZ & 3.02 & 3.14 & 3.03 & 3.15 & 3.19 & 3.09 & 3.03 & 3.10 \\
V7Z(d)\rlap{$^{a}$} & 2.78 & 3.71 & 2.72 & 3.36 & 3.04 & 3.11 & 3.08 & 2.87 \\
haVTZ & 3.28 & 3.57 & 3.24 & 3.37 & 2.77 & 3.32 & 3.70 & 3.38 \\
pwCVTZ\rlap{$^{b}$} & 3.05 & 3.27 & 3.04 & 3.37 & 3.80 & 3.41 & 3.24 & 3.24 \\
VQZ(f)\rlap{$^{a}$} & 3.39 & 3.85 & 3.32 & 3.71 & 3.53 & 3.76 & 3.88 & 3.64 \\
VQZ\rlap{$^{c}$} & 4.00 & 4.00 & 4.00 & 4.00 & 4.00 & 4.00 & 4.00 & 4.00 \\
V5Z(f) & 3.59 & 4.23 & 3.48 & 4.08 & 3.74 & 4.14 & 4.37 & 3.94 \\
haVQZ & 4.37 & 4.33 & 4.38 & 4.33 & 3.78 & 4.22 & 4.50 & 4.39 \\
V7Z(f) & 3.80 & 4.57 & 3.64 & 4.35 & 3.92 & 4.40 & 4.71 & 4.14 \\
pwCVQZ\rlap{$^{b}$} & 4.04 & 4.16 & 4.01 & 4.29 & 4.66 & 4.41 & 4.30 & 4.22 \\
V5Z(g) & 4.46 & 4.63 & 4.40 & 4.68 & 4.57 & 4.71 & 4.79 & 4.68 \\
V5Z\rlap{$^{c}$} & 5.00 & 5.00 & 5.00 & 5.00 & 5.00 & 5.00 & 5.00 & 5.00 \\
haV5Z & 5.42 & 5.40 & 5.43 & 5.39 & 4.93 & 5.27 & 5.47 & 5.51 \\
V7Z(g) & 4.93 & 5.30 & 4.81 & 5.31 & 5.06 & 5.34 & 5.52 & 5.24 \\
V6Z & 6.00 & 6.09 & 5.98 & 5.94 & 6.02 & 6.01 & 6.01 & 5.98 \\
\bottomrule
\end{tabular*}

\smallskip
\parbox{\textwidth}{\footnotesize
Model $E(b)=E_\infty+A\,(L_\textrm{eff}(b)+a)^{-3}$ per component, from a rank-1 fit of TAE basis-set deficits relative to REF1 (Riemann3 cc-pV\{6,7\}Z) over VTZ--V6Z, haVTZ--haV5Z, V5Z(f), V7Z(f), V5Z(g), V7Z(g); $n$ = species in the fit. Rows ordered by $L_\textrm{eff}$ for $[\textrm{CCSDT}]-[\textrm{CCSDT-3}]$. V$n$Z = cc-pV$n$Z; V$n$Z($l$): truncated at angular momentum $l$ (hydrogen one $l$ lower).
$^{a}$\,Not in the fit; projected onto the fitted species amplitudes.
$^{b}$\,Core--valence basis set, valence electrons correlated only; projected.
$^{c}$\,$L_\textrm{eff}=4$ and 5 by construction (anchors).}
\end{table*}

Table~\ref{tab:leff} presents the shifts and effective cardinal numbers for the various components based on the given data. Sextuple- and septuple-zeta extrapolation was used here as the reference. First of all, note that the optimized effective cardinal number for sextuple-zeta is 6.02 for CCSDT-3 $-$ CCSD(T), 6.01 for both CCSDT $-$ CCSDT-3 and CCSDT $-$ CCSD(T), and 5.98 for (Q), all satisfyingly close to the nominal cardinal number 6. Incidentally, the corresponding values for (T), for CCSDT correlation and for MP2 correlation are 5.94, 5.98, and 6.00, respectively.

So what \emph{are} the shifts? As a sanity check, we tested CC
SDT correlation and MP2 correlation, for which we found $a=-$0.04 and $+$0.45, respectively. For (T), we found $-$0.40. These shifts can be compared with the shifts in Ranasinghe and Petersson\cite{ranasinghe2013ccsd} for a different sequence of basis sets: $+1/4$ for MP2, $-2/3$ for (T). These authors also asymptotically found $+3/2$ for CCSD $-$ MP2, compared to $+1.87$ presently. While no direct comparison is possible --- those authors included an $L^{-5}$ term in their expansion, plus a somewhat murky `interference correction' for CCSD $-$ MP2 --- we are in qualitative agreement. (Note that positive vs. negative $a$ means locally slower vs. faster than $L^{-3}$.) Karton's exponents from Table V of Ref.\cite{Karton2020} correspond for CCSDT $-$ CCSD(T) to $a$ shifts \{0.25,0.72,0.43\} for \{\{D,T\},\{T,Q\},\{Q,5\}\} basis set pairs; presently we obtain +0.69. For (Q), his exponents correspond to $a$=\{0.00,-0.39,-0.55\}, respectively, while we have $-0.66$. Again, qualitative agreement prevails.

For CCSDT-3 $-$ CCSD(T), our shift is plus 1.1. For CCSDT $-$ CCSDT-3, it is plus 0.84, and for the sum of the two, plus 0.69, where positive shifts correspond to a convergence that is slower than nominal cardinal number to the inverse third power. But ultimately, asymptotically, in the very large basis set limit, this Petersson formula will reach $L^{-3}$. In contrast, for (Q), we obtain a shift, albeit with fewer data points, only 32 instead of 42, of $-$0.65, which indicates faster convergence than $L^{-3}$.

Now we can look upon these effective cardinal numbers in several ways. First of all, you can  consider those for the cc-pVTZ basis set and see to what extent they differ from the nominal 3. The answer is: significantly so, 3.19, 3.09, 3.03, and 3.10 for the four series, respectively.

Another insight we can obtain from these numbers is the quality of truncated basis sets. For example, while the septuple-zeta truncated at d functions has a nominal cardinal number of two because it stops at d functions, its cardinal numbers for the four series reach 3.04, 3.11, 3.08, and 2.87, respectively. Two of these are actually higher than for cc-pVTZ, and essentially all of them a full unit higher than for cc-pVDZ.
(We note in passing that Karton (Table III in Ref.\cite{Karton2018PostCCSDT}) previously found truncated cc-pVTZ to be distinctly superior to cc-pVDZ.) 

For example, septuple-zeta truncated at f functions is at 3.92, 4.40, 4.71, and 4.14, respectively. All except one are higher or considerably higher than even the cc-pVQZ, nominal 4. And in fact, the same is true even of the quintuple-zeta truncated at f. While the quadruple-zeta truncated at f clocks in at 3.53, 3.76, 3.88, and 3.64, getting pretty close to the fixed values of 4 for the cc-pVQZ basis set. This is not a behavior one sees, for example, for the MP2 correlation energy, 3.39, or for the CCSD correlation energy, 3.32, which is closer to the cc-pVTZ values, which top out at the same angular momentum.

%In the Supporting Information, one can find the same extrapolation recast in Schwenke format, in $L^{-\alpha}$ exponent format, and in Petersson shift $(L+a)^{-3}$ format.

In Table~\ref{tab:statspost}, the reader can find the RMSD in \kcalmol of the two-point extrapolation for this particular series with this particular basis set pair. cc-pVDZ and cc-pVTZ is only satisfactory for CCSDT $-$ CCSD(T), really. For (Q) it does more harm than good, as was established by Karton et al. in the W4 paper\cite{jmlm200} and in the HEAT-3 paper\cite{HEAT3} That in fact, for polar molecules, (Q) starting at double-zeta is non-monotonic. Double-zeta can be larger than triple-zeta, and hence extrapolation from double and triple gives a disastrous result. In W4 theory, we applied for that reason a fixed scaling factor 1.10; here we find this causes RMSD=0.04 \kcalmol, about one-half the error of (Q)/V\{D,T\}Z extrapolation. The latter is admittedly still superior to raw pVTZ numbers, but that is `damning with faint praise'. 

Triple- and quadruple-zeta, however, as seen in the table, can work remarkably well, even for CCSDT $-$ CCSDT-3, 0.02 \kcalmol RMS, 0.007 \kcalmol for (Q), 0.01 for CCSDT $-$ CCSD(T). In the latter, apparently error compensation takes place between its two components.

For comparison, using simply the largest basis set available incurs errors over the four series of 0.05, 0.14, 0.09, and 0.05 \kcalmol. Even going from quadruple- to quintuple-zeta basis sets, the largest basis set still has RMS errors of 0.08 \kcalmol for CCSDT $-$ CCSDT-3, 0.05 for CCSDT $-$ CCSD(T). In contrast, with extrapolation, we're down to 0.06, 0.01, 0.09, and 0.02, respectively.

% preamble: \usepackage{booktabs,siunitx,threeparttable}
\begin{table*}[!t]
\footnotesize \setlength{\tabcolsep}{4pt}
  \begin{threeparttable}
  \caption{Errors (\kcalmol) of the post-CCSD(T) contributions to the TAE of some representative second-row species, relative to the cc-pV\{T,Q\}Z limit $[\mathrm{CCSDT}-\mathrm{CCSD(T)}]$ from \{D,T\} extrapolation with $\alpha=3$; $[(\mathrm{Q})]$ as $1.10\times$ the TZ value. ``Diff'' is the plain minus the +d sum error. 
  RMSD statistics refer to the 65 second-row species in W4-17.
%  The 18 species are the union of the ten largest $|\text{sum error}|$ (plain) and the ten largest $|\text{diff}|$.
  }
  \label{tab:dt_plain_vs_d}
  \sisetup{table-format=+1.3, retain-explicit-plus}
  \begin{tabular}{l S S S S S S S}
    \toprule
    & \multicolumn{2}{c}{$[\mathrm{CCSDT}-\mathrm{CCSD(T)}]$}
    & \multicolumn{2}{c}{$[(\mathrm{Q})]$}
    & \multicolumn{2}{c}{CCSDT(Q)$-$CCSD(T) total\tnote{a}} & {diff} \\
    \cmidrule(lr){2-3}\cmidrule(lr){4-5}\cmidrule(lr){6-7}
    Species & {V\{D,T\}Z} & {V(\{D,T\}+d)Z} & {VTZ} & {V(T+d)Z} & {V\{D,T\}Z} & {V(\{D,T\}+d)Z} & \\
    \midrule
    HClO$_4^{*}$  & +0.183 & +0.124 & -0.107 & -0.159 & +0.076 & -0.035 & +0.110 \\
    SO$_3$        & +0.112 & +0.081 & +0.016 & -0.027 & +0.128 & +0.055 & +0.073 \\
    SF$_6$        & +0.059 & +0.031 & -0.031 & -0.070 & +0.028 & -0.039 & +0.067 \\
    HOClO$_2^{*}$ & +0.164 & +0.127 & -0.098 & -0.126 & +0.065 & +0.001 & +0.064 \\
    SO$_2$        & +0.111 & +0.089 & +0.029 & -0.006 & +0.139 & +0.083 & +0.056 \\
    PF$_5$        & -0.029 & -0.048 & -0.035 & -0.051 & -0.064 & -0.099 & +0.035 \\
    S$_2$O        & +0.119 & +0.094 & -0.036 & -0.044 & +0.083 & +0.050 & +0.033 \\
    CS$_2$        & +0.049 & +0.036 & -0.112 & -0.069 & -0.063 & -0.032 & -0.030 \\
    ClF$_3^{*}$   & +0.087 & +0.077 & -0.107 & -0.118 & -0.020 & -0.041 & +0.021 \\
    PF$_3^{*}$    & +0.027 & +0.013 & -0.054 & -0.061 & -0.027 & -0.048 & +0.021 \\
    %OClO          & +0.097 & +0.079 & -0.027 & -0.029 & +0.070 & +0.050 & +0.021 \\
    SiO           & +0.039 & +0.036 & +0.058 & +0.047 & +0.097 & +0.083 & +0.014 \\
    Si$_2$H$_6$   & -0.169 & -0.177 & -0.028 & -0.030 & -0.197 & -0.207 & +0.011 \\
    P$_2$         & -0.033 & -0.033 & -0.080 & -0.070 & -0.113 & -0.103 & -0.011 \\
    PH$_3$        & -0.083 & -0.089 & -0.013 & -0.014 & -0.096 & -0.103 & +0.007 \\
    P$_4$         & -0.161 & -0.184 & -0.205 & -0.174 & -0.365 & -0.358 & -0.007 \\
    SiH$_4$       & -0.093 & -0.098 & -0.006 & -0.007 & -0.099 & -0.105 & +0.006 \\
    S$_4$ ($C_{2v}$) & +0.300 & +0.230 & -0.139 & -0.073 & +0.161 & +0.157 & +0.004 \\
    ClOOCl$^{*}$  & +0.033 & +0.015 & -0.144 & -0.125 & -0.111 & -0.110 & -0.001 \\
    \midrule
    RMSD (66 species) & {} & {} & 0.082 & 0.069 & 0.075 & 0.070 & 0.024 \\
    \bottomrule
  \end{tabular}
  \begin{tablenotes}\footnotesize
    \item[a] The cc-pV\{T,Q\}Z and cc-pV(\{T,Q\}+d)Z limits themselves differ by an RMS of only 0.0034 for $[\mathrm{CCSDT}-\mathrm{CCSD(T)}]$ (45 species), 0.0063 for $[(\mathrm{Q})]$ and 0.0057 for the sum (35 species each); the largest deviations are 0.013 and 0.019 (S$_4$) and 0.012 (S$_2$). ClOO, with its erraticaly bifurcating UHF solution, is excluded.\\
    \item[*] From  cc-pVTZ and cc-pV\{T,Q(f)\}Z).
  \end{tablenotes}
  \end{threeparttable}
\end{table*}

What about diffuse-function-augmented basis sets? Somewhat peculiarly, the cardinal numbers for CCSDT-3 $-$ CCSD(T) are actually somewhat smaller for heavy-augmented basis sets than for the unaugmented ones. For CCSDT $-$ CCSDT-3, as well as for CCSDT $-$ CCSD(T), the opposite is true, however. And likewise, for (Q), heavy augmentation always gives an improvement from 3.10 to 3.38 for triple-zeta, from 4 to 4.39 for quadruple-zeta, and from 5 to 5.51 for quintuple-zeta.

This kind of prompts the question: how does augmentation compare to a truncated version of the next basis set up, which in fact is the same size? Here the answer is fairly unequivocal. From cc-pVQZ to cc-pV5Z truncated at g, for example, we go in the four series from 4 to 4.57, 4.71, 4.79, and 4.68, respectively.

In contrast, for the MP2 energy, the heavy-augmented comes in at 4.37, and the quintuple-zeta truncated at $g$, very close to that at 4.46. Behavior for (T) is actually similar to the higher order in that the heavy-augmented clocks in at 4.33, and the truncated quintuple-zeta at 4.68.

A similar comparison for triple-zeta would be with heavy-augmented triple-zeta and with quadruple-zeta truncated at f. Unequivocally, again quadruple-zeta truncated at f wins it for the post-CCSD(T) contributions. For the MP2 correlation energy, the two values are actually quite close, whereas for (T), once again the behavior is very similar to the post-CCSD(T).

%\subsection{A brief aside on cc-pV(n+d)Z basis sets}
\subsection{A brief aside on cc-pV($n$+d)Z basis sets}

%It is well known\cite{Bauschlicher1995,jmlm107,jmlm121,Wilson2001plusd,jmlm191} that second-row elements in high oxidation states require the addition of a `tight' (i.e., high-exponent) $d$ function to the basis set to describe the inner part of the vacant $3d$ orbital, which in high oxidation states drops low enough that it can accept back-bonding from O and F lone pairs.\cite{jmlm191} It is also well known that the pre-eminent effect is a drastic (as high as 40-50 \kcalmol for \ce{SO3} or \ce{HClO4} in DZ basis sets) increase in the SCF component of the atomization energy.

It is well known\cite{Bauschlicher1995,jmlm107,jmlm121,Wilson2001plusd,jmlm191} that second-row elements in high oxidation states require the addition of a `tight' (i.e., high-exponent) $d$ function to the basis set to describe the inner part of the vacant $3d$ orbital, which in high oxidation states drops low enough that it can accept back-bonding from O and F lone pairs.\cite{jmlm191} It is also well known that the pre-eminent effect is a drastic increase in the SCF component of the atomization energy: in DZ basis sets, as much as 40 \kcalmol for \ce{HClO4} and 29 \kcalmol for \ce{SO3}.

%But is there any impact at all for post-CCSD(T) Karton\cite{Karton2022Tightening} studied the issue in great detail, and found that generally the changes from cc-pVDZ and cc-pVTZ to cc-pV(D+d)Z and cc-pV(T+d)Z tend to nearly cancel between CCSDT $-$ CCSD(T) and (Q). This agrees with our own results; in addition, we found a negligible RMSD=0.006 \kcalmol  between cc-pV\{T,Q\}Z and cc-pV(\{T,Q\}+d)Z. In Table~\ref{tab:dt_plain_vs_d} one can see that the differences between cc-pV\{D,T\}Z and cc-pV(\{D,T\}+d)Z can reach as high as 0.11, 0.07, 0.06, and 0.06 \kcalmol, respectively, for 0.11 \kcalmol for \ce{HClO4}, 0.07 for \ce{SO3} and \ce{SF6}, and 0.06 \kcalmol for \ce{HOClO2} and \ce{SO2}, and that cc-pV(\{D,T\}+d)Z tends to be somewhat closer to the basis set limit than pV\{D,T\}Z (although \ce{PF5} and \ce{SF6} buck that trend). At the same time, for species like \ce{P4}, \ce{S4}, and \ce{Si2H6}, both \{D,T\} options make similar and unacceptable errors (0.36, 0.16, 0.20 kcal/mol), and the only real answer is pressing on to a VQZ option. As already noted repeatedly,\cite{jmlm330,jmlm340} cc-pVQZ(f) constitutes a viable fallback option if the resource requirements for cc-pVQZ become insurmountable.

But is there any impact at all on the post-CCSD(T) contributions? Karton\cite{Karton2022Tightening} studied the issue in great detail, and found that the changes from cc-pVDZ and cc-pVTZ to cc-pV(D+d)Z and cc-pV(T+d)Z tend to nearly cancel between CCSDT $-$ CCSD(T) and (Q). Our own results largely agree: for most species, the two changes have opposite signs. The exceptions, however, are precisely the `pseudohypervalent'\cite{Cioslowski1993,ReedSchleyer1990,jmlm215} species: for \ce{HClO4}, \ce{SO3}, \ce{SF6}, and \ce{SO2}, the tight-$d$ effects on CCSDT $-$ CCSD(T) and (Q) have the same sign and add up. Close to the basis set limit, the issue disappears: the cc-pV\{T,Q\}Z and cc-pV(\{T,Q\}+d)Z limits differ by an RMS of only 0.003 and 0.006 \kcalmol for CCSDT $-$ CCSD(T) and (Q), respectively.

With \{D,T\} extrapolation for CCSDT $-$ CCSD(T) and (Q) scaled as $1.10\times$ the cc-pVTZ value (Table~\ref{tab:dt_plain_vs_d}), the cc-pV\{D,T\}Z and cc-pV(\{D,T\}+d)Z sums differ by as much as 0.11, 0.07, 0.07, 0.06, and 0.06 \kcalmol for \ce{HClO4}, \ce{SO3}, \ce{SF6}, \ce{HOClO2}, and \ce{SO2}, respectively. For the oxides and oxyacids, cc-pV(\{D,T\}+d)Z tends to be closer to the basis set limit than cc-pV\{D,T\}Z; for the fluorides (\ce{PF5}, \ce{SF6}, \ce{ClF3}, \ce{PF3}) the reverse is true. Over all 66 second-row species, the two variants perform almost identically (RMSD 0.070 vs.\ 0.075 \kcalmol). The verdict also depends on the treatment of (Q): if (Q) is likewise obtained by \{D,T\} extrapolation, cc-pV(\{D,T\}+d)Z is the poorer of the two for most of these species. Meanwhile, for species like \ce{P4}, \ce{S4}, and \ce{Si2H6}, both \{D,T\} options make similar and unacceptable errors (0.36, 0.16, and 0.20 \kcalmol, respectively), and the only real answer is pressing on to a VQZ option. As already noted repeatedly,\cite{jmlm330,jmlm340} cc-pVQZ(f) constitutes a viable fallback if the resource requirements for cc-pVQZ become insurmountable: here, cc-pV\{T,Q(f,d)\}Z reproduces the cc-pV\{T,Q\}Z limits for the second-row species to within an RMSD of 0.04 and 0.025 \kcalmol for CCSDT $-$ CCSD(T) and (Q), respectively.

\section{\label{sec:conclusions}Conclusions}

We were able to obtain higher-order triples corrections for about four dozen molecules with basis sets as large as cc-pV7Z, which are so close to the one-particle basis set limit that the details of extrapolation no longer are thermodynamically significant. We were also able to do this for connected quadruples through cc-pV6Z (and for a subset through cc-pV7Z). We are able to draw the following conclusions:
\begin{enumerate}
    \item for pVQZ and larger basis sets, basis set convergence behavior is well-described by the Petersson expression $(L+a)^{-3}$, where the shift $a$ equals +1.10 for CCSDT-3 $-$ CCSD(T), +0.84 for CCSDT $-$ CCSDT-3, +0.69 for the sum of the two, and $-0.66$ for (Q);
    \item convergence behavior of the higher-order triples contribution CCSDT $-$ CCSD(T) is a superposition of opposite behaviors in the two components: a shallow upward slope with $L$ for CCSDT-3 $-$ CCSD(T) and a steeper downward slope for CCSDT $-$ CCSDT-3. 
    The latter is itself a superposition of fast convergence for higher-order triples coupling terms that make of CCSDT - CCSDT-4, and very slow convergence of the `triples-on-triples' ladder term CCSDT-4 - CCSDT-3 with the leading fifth-order term in perturbation theory with $E^{(5)}_{TT}$, and which
    dominates in the large basis set limit;
    \item CCSDT $-$ CCSDT-3 appears to be responsible for the at times surprisingly slow convergence behavior of $T_3-(T)$ in certain molecules;
    \item radial flexibility is at least as important as angular flexibility (more angular momenta) for basis set convergence of (T) and especially post-(T) components. This contrasts with the MP2 and CCSD correlation energies, where angular momentum is paramount;
%    \item CCSDT-3 $-$ CCSD(T)  and CCSDT $-$ CCSDT-3 are anticorrelated (in the Pearson sense of the word) 
    \item basis set convergence in CCSDT $-$ CCSDT-3 is almost invariably monotonic (except for the very small values in some hydrides), while in CCSDT-3 $-$ CCSD(T) the smallest pVDZ basis set breaks monotonicity. In \{D,T\} extrapolations, this may lead to erratic behavior: usually this cancels between CCSDT-3 $-$ CCSD(T) and CCSDT $-$ CCSDT-3, but for Si and P compounds (e.g., for \ce{P4}) this breaks down;
    \item in contrast to the higher-order triples, basis set convergence of (Q) is surprisingly smooth beyond cc-pVTZ, and cc-pV\{5,6\}Z values can be regarded as reference quality;
    \item we propose extrapolation parameters or all of these, as well as `effective cardinal numbers' for basis set sequences other than cc-pVnZ;
    \item the somewhat counterintuitive behavior of saturated fluorides and chlorides found in the W4-17 paper\cite{jmlm273} and discussed by Karton\cite{KartonARCCChapter} can be traced to strongly negative CCSDT$-$CCSDT-3 in these species; for the fluorides, CCSDT-3$-$CCSD(T) have small positive contributions, which turn negative themselves for the chlorides;
    \item post-CCSDT(Q) contributions converge rapidly with the basis set, and benefit more from radial than from angular basis set expansion. Both features reflect the increasingly static correlation character of these contributions.
\end{enumerate}

%As a final observation,[insert]

\begin{acknowledgement}

\noindent This work was supported by the Minerva Foundation, Munich, Germany.  
JMLM acknowledges the sabbatical hospitality of the Quantum Theory Project, University of Florida, and is grateful for several helpful discussions with its late lamented chair, Prof. John F. Stanton (1961-2025) and with the `Stantonites', particularly Drs. James H. Thorpe, Gregory H. Jones, and Peter R. Franke, as well as Ms. Kaila E. Weflen.

CHEMFARM is supported by the Ben May Center for Chemical Theory and Computation, Weizmann Institute of Science.
\end{acknowledgement}

\begin{suppinfo}

\begin{itemize}
\item Python scripts used for collating and processing the CFOUR output data
\item Total energies in CSV (comma-separated values) format.
\item Text files with summaries of thermochemical values
\end{itemize}

\end{suppinfo}

\subsection*{Data and Software Availability}
The data underlying this study are available in the published article and the Supporting Information. Further details may be obtained upon reasonable request.

\begingroup
\renewcommand{\bibfont}{\fontsize{9pt}{10.5pt}\selectfont}
\setlength{\bibsep}{0pt}
\bibliography{postCCSDtExtrap}
\endgroup

\end{document}